\documentclass[aps,prl,twocolumn,floatfix,letterpaper,longbibliography]{revtex4-2}
\usepackage{graphicx}
\usepackage{bm}
\usepackage{amsmath,amsfonts}
\usepackage{amsbsy}
\usepackage[colorlinks=true,linkcolor=blue,urlcolor=blue,citecolor=blue,breaklinks=true]{hyperref}
\usepackage[utf8]{inputenc}
\DeclareUnicodeCharacter{03BC}{\mbox{$\mu$}}
\DeclareUnicodeCharacter{2009}{ }

\begin{document}

\title{Sound in a system of chiral one-dimensional fermions}

\author{K. A. Matveev}

\affiliation{Materials Science Division, Argonne National Laboratory,
  Argonne, Illinois 60439, USA}

\date{\today}

\begin{abstract}

  We consider a system of one-dimensional fermions moving in one
  direction, such as electrons at the edge of a quantum Hall system.
  At sufficiently long time scales the system is brought to
  equilibrium by weak interactions between the particles, which
  conserve their total number, energy, and momentum.  Time evolution
  of the system near equilibrium is described by hydrodynamics based
  on the three conservation laws.  We find that the system supports
  three sound modes.  In the low temperature limit one mode is a pure
  oscillation of particle density, analogous to the ordinary sound.
  The other two modes involve oscillations of both particle and
  entropy densities.  In the presence of disorder, the first sound
  mode is strongly damped at frequencies below the momentum relaxation
  rate, whereas the other two modes remain weakly damped.

\end{abstract}
\maketitle

Small perturbations of density propagate through ordinary fluids in
the form of sound waves \cite{landau_fluid_2013}.  Oscillations of
density in a sound wave are accompanied by oscillations of entropy
density, such that the entropy per particle retains its equilibrium
value.  An important exception to the above picture is superfluid
$^4$He, whose quantum nature results in the existence of two sound
modes.  The first sound is predominantly a wave of density, similar to
sound in ordinary fluids, whereas the second sound is a wave of
entropy \cite{landau_fluid_2013, khalatnikov_introduction_2000}.

It was recently shown that the superfluid behavior characterized by
the existence of two sound modes is a generic feature of the
one-dimensional quantum fluids \cite{matveev_second_2017,
  matveev_hybrid_2018}.  This is a consequence of the fact that in
addition to the number of particles, energy, and momentum, the
collisions of particles in these systems conserve an additional
quantity $J$.  The latter has the meaning of the difference of the
numbers of the right- and left-moving particles.  Although the
scattering processes changing $J$ are not strictly forbidden, their
rate is exponentially small at low temperatures,
$\tau^{-1}\propto e^{-D/T}$ \cite{lunde_three-particle_2007,
  micklitz_transport_2010, matveev_equilibration_2012}.  (Here $D$ is
the bandwidth of the system and $T$ is the temperature.)  At low
frequencies $\omega\ll\tau^{-1}$ the conservation of $J$ is violated,
and the system behaves as an ordinary fluid, with a single sound mode.
On the other hand, in a broad range of frequencies between $\tau^{-1}$
and the much larger rate of quasiparticle relaxation
$\tau_{\rm ex}^{-1}$, the additional conservation law changes the
dynamics of the quantum fluid dramatically, resulting in two sound
modes.

In this paper we study sound in a gas of chiral one-dimensional
fermions.  The best known example of such a system is the edge of the
two-dimensional electron gas in the integer quantum Hall regime
\cite{halperin_quantized_1982}.  The chiral nature of transport in
this system was demonstrated by observing propagation of the density
pulses along the edge \cite{ashoori_edge_1992}.  More recently, an
analog of the quantum Hall system was created with cold atoms, and the
chiral nature of the edge states has also been demonstrated
experimentally \cite{goldman_direct_2013}.  We show below that the
number of sound modes in a chiral system is equal to the number of
conserved quantities.
% (In nonchiral systems the number of sound modes
% cannot exceed half the number of conservation laws.)
In the absence of disorder, collisions between fermions conserve the
number of particles, energy, and momentum, resulting in three sound
modes.  By breaking the conservation of momentum, static disorder
reduces the number of sound modes to two.  This is in contrast to
nonchiral systems, where no sound may propagate in the presence of
disorder.

We focus on the simplest case of a single-component chiral Fermi gas,
which corresponds to occupation fraction $\nu=1$ in the quantum Hall
realizations of the system.  At low temperatures, the energy spectrum
of fermions, which we assume to be spinless, can be expanded in
momentum near the Fermi point,
\begin{equation}
  \label{eq:energy_spectrum}
  \epsilon_p=v_F^{}p + \frac{p^2}{2m}+\frac{\lambda p^3}{2m^2v_F}+\ldots.
\end{equation}
Here we measure the momentum $p$ from the Fermi point, $v_F$ is the
Fermi velocity, parameter $m$ has the dimension of mass, while
$\lambda$ is dimensionless.  For simplicity, we assume that the
interactions between particles are weak and account for them only to
the extent that they bring the system to thermal equilibrium, with the
characteristic relaxation rate $\tau_{\rm ex}^{-1}$.  Stronger
interactions would alter our results for the sound velocities but not
the fundamental features, such as the number of the sound modes.

We note that it is important to properly account for the nonlinear
corrections to the energy spectrum in Eq.~(\ref{eq:energy_spectrum}).
Indeed, for the linear spectrum the energy of the system is determined
by its momentum, $E=v_F P$, and the two corresponding conservation
laws become equivalent.  Because of that, the subsequent calculations
simplify considerably if instead of conservation of momentum one
discusses conservation of the quantity $\Theta=E-v_FP$.
% Because of that, instead of
% conservation of momentum, it is convenient to discuss conservation of
% the quantity $\Theta=E-v_FP$.
In this approach, the equilibrium state
of the system is fully determined by specifying the total number of
particles $N$, energy $E$, and $\Theta$.  Correspondingly, the
occupation numbers of the fermionic states
\begin{equation}
  \label{eq:Fermi_distribution_theta}
  f_p=\frac{1}{e^{(\epsilon_p-\mu-\gamma\theta_p)/T}+1},
  \quad
  \theta_p=\epsilon_p-v_Fp,
\end{equation}
are controlled by three parameters: the chemical potential $\mu$, the
temperature $T$, and the dimensionless parameter $\gamma$.  The values
of $\mu$, $T$, and $\gamma$ are determined by the magnitudes of the
conserved quantities $N$, $E$, and $\Theta$.

We are interested in the dynamics of the system at frequencies
$\omega$ well below the quasiparticle relaxation rate
$\tau_{\rm ex}^{-1}$.  In this regime, the Fermi-Dirac form
(\ref{eq:Fermi_distribution_theta}) of the occupation numbers applies
at every point in space, but the parameters $\mu$, $T$, and $\gamma$
may depend on the position $x$ and time $t$.  The dynamics of the
system is fully described by the continuity equations expressing the
three conservation laws,
\begin{subequations}
  \begin{eqnarray}
    \label{eq:continuity_number}
    \partial_t n +\partial_x j_n&=&0,
\\
    \label{eq:continuity_energy}
    \partial_t \varepsilon +\partial_x j_\varepsilon&=&0,
\\
  \label{eq:continuity_theta}
  \partial_t\vartheta+\partial_xj_\vartheta&=&0.
  \end{eqnarray}
  \label{eq:continuity_equations}%
\end{subequations}
Here $n$, $\varepsilon$, and $\vartheta$ are the densities of $N$,
$E$, and $\Theta$, respectively, whereas $j_n$, $j_\varepsilon$, and
$j_\vartheta$ are the corresponding currents.  For noninteracting
fermions they are given by
\begin{equation}
  \label{eq:densities}
  \begin{array}[c]{lll}
    \displaystyle
    n=\int \frac{dp}{h} f_p,
    & \displaystyle\varepsilon=\int \frac{dp}{h} \epsilon_p f_p,
    & \displaystyle\vartheta=\int \frac{dp}{h} \theta_p f_p,
    \\[2ex]
    \displaystyle j_n=\int \frac{dp}{h}v_p f_p,
    &\displaystyle j_\varepsilon=\int \frac{dp}{h}v_p\epsilon_p f_p,
    &    \displaystyle j_\vartheta=\int \frac{dp}{h}v_p\theta_p f_p.
  \end{array}
\end{equation}
where $h$ is the Planck's constant and $v_p=\partial_p\epsilon_p$ is
the velocity of the fermion with momentum $p$.  Substitution of the
densities and currents defined by Eq.~(\ref{eq:densities}) with $f_p$
given by Eq.~(\ref{eq:Fermi_distribution_theta}) into the continuity
equations (\ref{eq:continuity_equations}) gives three equations upon
the three parameters $\mu(x,t)$, $T(x,t)$, and $\gamma(x,t)$ of the
distribution function.

To obtain sound modes of the system, we assume that these parameters
oscillate according to
\begin{eqnarray}
  \label{eq:plain_wave}
  &&\mu(x,t)=\delta\mu\, e^{i(qx-\omega t)},
  \quad
  T(x,t)=T+\delta T\, e^{i(qx-\omega t)},
  \nonumber\\
  &&\gamma(x,t)=\delta\gamma\, e^{i(qx-\omega t)},
\end{eqnarray}
with small amplitudes $\delta\mu$, $\delta T$, and $\delta\gamma$ and
linearize Eq.~(\ref{eq:continuity_equations}) in these parameters.
This yields a system of three linear equations, which we write in the
matrix form
\begin{equation}
  \label{eq:eigenvalue_problem}
  \left(\omega \widehat D - q \widehat J\right)\Psi = 0.
\end{equation}
Here
\begin{equation}
  \label{eq:matrices}
  \widehat D=\!
  \left(\!
  \begin{array}[c]{ccc}
    \partial_\mu n & \partial_T n & \partial_\gamma n\\[1ex]
    \partial_\mu \varepsilon & \partial_T \varepsilon
   &\partial_\gamma \varepsilon\\[1ex]
     \partial_\mu \vartheta & \partial_T \vartheta
   &\partial_\gamma \vartheta
  \end{array}
  \!\right)\!,
  \ \ 
  \widehat J=\!
  \left(\!
  \begin{array}[c]{ccc}
    \partial_\mu j_n & \partial_T j_n & \partial_\gamma j_n\\[1ex]
    \partial_\mu j_\varepsilon & \partial_T j_\varepsilon
   &\partial_\gamma j_\varepsilon\\[1ex]
     \partial_\mu j_\vartheta & \partial_T j_\vartheta
   &\partial_\gamma j_\vartheta
  \end{array}
  \!\right),
\end{equation}
and the column vector $\Psi=(\delta\mu,\delta T,\delta\gamma)^{\rm
  T}$.  The sound modes are given by nonvanishing solutions of
Eq.~(\ref{eq:eigenvalue_problem}), which exist only when $\omega$ and
$q$ satisfy the condition
\begin{equation}
  \label{eq:determinant_condition}
  \det\left(\omega \widehat D - q \widehat J\right)=0.
\end{equation}
For a given $q$, the above condition is a cubic equation for $\omega$,
which in general has three solutions.  This conclusion applies to any
chiral system with conserved number of particles, energy, and
momentum.

In the particular case of the gas of chiral fermions at low
temperature, further progress can be made by evaluating the matrices
(\ref{eq:matrices}) to leading order in $T/mv_F^2$ using
Eqs.~(\ref{eq:densities}) and (\ref{eq:Fermi_distribution_theta}).  We
obtain
\begin{equation}
  \label{eq:D}
  \widehat D=
  \!\left(\!
\begin{array}{ccc}
  \displaystyle\frac{1}{h {v_F}}
  &\displaystyle -\frac{\pi ^2 T}{3 h m v_F^3}
  &\displaystyle \frac{\pi ^2 T^2}{6 h m v_F^3} \\[3ex]
  \displaystyle-\frac{\pi ^2 T^2}{3 h m v_F^3}
  &\displaystyle \frac{\pi ^2 T}{3 h v_F}
  &\displaystyle \frac{7 \pi ^4 (\lambda -2) T^4}{30 h m^2 v_F^5} \\[3ex]
  \displaystyle\frac{\pi ^2 T^2}{6 h m v_F^3}
  &\displaystyle \frac{7 \pi ^4 (\lambda -2) T^3}{30 h m^2 v_F^5}
  &\displaystyle \frac{7 \pi ^4 T^4}{60 h m^2 v_F^5}
\end{array}
\!\right)\!
\end{equation}
and
\begin{equation}
  \label{eq:J}
  \widehat J=
  \!\left(\!
\begin{array}{ccc}
  \displaystyle\frac{1}{h}
  & 0
  &\displaystyle \frac{\pi ^2 T^2}{6 h m v_F^2} \\[3ex]
  0
  &\displaystyle \frac{\pi ^2 T}{3 h}
  &\displaystyle \frac{7 \pi^4 (\lambda -1) T^4}{30 h m^2 v_F^4}
   \\[3ex]
  \displaystyle\frac{\pi ^2 T^2}{6 h m v_F^2}
  &\displaystyle \frac{7 \pi ^4 (\lambda -1) T^3}{30 h m^2 v_F^4}
  &\displaystyle \frac{7 \pi ^4 T^4}{60 h m^2 v_F^4}
\end{array}
\!\right).
\end{equation}
Then by solving Eq.~(\ref{eq:determinant_condition}) we find three
values of the sound velocity $\omega/q$,
\begin{equation}
  \label{eq:velocities_3}
  v_1=v_F,
  \quad
  v_\pm=v_F\left(1\pm\sqrt{\frac75}\frac{\pi T}{mv_F^2}\right).
\end{equation}
where we kept only the terms up to first order in $T/mv_F^2$.  In this
approximation the relative magnitudes of the oscillations of $\mu$,
$T$, and $\gamma$ are given by
\begin{equation}
  \label{eq:deltas_3_0}
  \delta T = 0,
  \quad
  \frac{\delta\mu}{\delta\gamma}=-\frac{7}{10}\frac{\pi^2T^2}{mv_F^2},
\end{equation}
for the mode propagating at velocity $v_1=v_F$ and
\begin{equation}
  \label{eq:deltas_3_pm}
  \delta\mu=0,
  \quad
  \frac{\delta T}{\delta\gamma}=\pm\sqrt{\frac{7}{20}}\frac{\pi T^2}{mv_F^2},
\end{equation}
for the solutions with sound velocities $v_\pm$.  One can use the
expression (\ref{eq:D}) for the matrix $\widehat D$ to obtain the
relative magnitude of the oscillations of density $\delta n$ and
entropy density $\delta s=\delta\varepsilon/T$.  At $T/mv_F^2\to0$ one
finds $\delta s/\delta n=0$ for the solution with velocity $v_1=v_F$
and $\delta s/\delta n=\pm\pi\sqrt{7/5}$ for the modes with velocities
$v_\pm$.

Let us now discuss the effect of static disorder on the sound modes.
Disorder breaks the translation invariance of the system, and the
total momentum is no longer a conserved quantity.  This brings about
two changes in the above discussion.  First, the parameter $\gamma$ in
the Fermi-Dirac distribution (\ref{eq:Fermi_distribution_theta}) of
the occupation numbers is now permanently set to zero.  Second, only
the continuity equations (\ref{eq:continuity_number}) and
(\ref{eq:continuity_energy}) are satisfied.  Upon linearization in
$\delta\mu$ and $\delta T$, Eq.~(\ref{eq:eigenvalue_problem}) still
applies, but the matrices $\widehat D$ and $\widehat J$ are given by
the first two rows and columns of the corresponding expressions in
Eq.~(\ref{eq:matrices}), and $\Psi=(\delta\mu, \delta T)^{\rm T}$. At
a given $q$, the condition (\ref{eq:determinant_condition}) is now a
quadratic equation for $\omega$.  It has two solutions corresponding
to two sound modes.  For the chiral Fermi gas at low temperatures we
solve Eq.~(\ref{eq:determinant_condition}) using the matrices
$\widehat D$ and $\widehat J$ given by the first two rows and columns
of Eqs.~(\ref{eq:D}) and (\ref{eq:J}).  The resulting sound velocity
$\omega/q$ takes two possible values
\begin{equation}
  \label{eq:velocities_2}
  v_\pm=v_F\left(1\pm\frac{\pi}{\sqrt3}\frac{T}{mv_F^2}\right).
\end{equation}
Here again we omitted the terms of second and higher powers in $T$.
The eigenvectors $\Psi$ corresponding to these two solutions have
$\delta \mu/\delta T=\pm\pi/\sqrt3$.  The corresponding oscillations
of density $\delta n$ and entropy density $\delta s$ in the two sound
modes satisfy the conditions $\delta s/\delta n=\pm \pi/\sqrt3$.

To gain further insight into the effect of disorder on the sound
modes, we consider the case of weak disorder.  The latter condition
means that the rate of relaxation of momentum $\tau_P^{-1}$ due to
disorder is small compared to the quasiparticle relaxation rate
$\tau_{\rm ex}^{-1}$.  At $\omega\ll\tau_{\rm ex}^{-1}$ the
quasiparticles of the Fermi gas are in equilibrium, and the occupation
numbers take the form (\ref{eq:Fermi_distribution_theta}), but the
parameter $\gamma$ experiences relaxation due to the momentum
nonconserving scattering processes.  Assuming this relaxation takes
the usual form
\begin{equation}
  \label{eq:dot_gamma}
  \dot\gamma=-\frac{\gamma}{\tau_P^{}},
\end{equation}
we now obtain the effect of these processes on the three sound modes.

First, we note that a nonzero $\dot\gamma$ is accompanied by nonzero
$\dot\mu$ and $\dot T$.  Indeed, the remaining laws of conservation of
the particle number and energy imply $\dot n=0$ and
$\dot\varepsilon=0$.  The set of three quantities
$\dot\Phi=(\dot n,\dot\varepsilon, \dot\vartheta)^{\rm T}$ is obtained
from $\dot\Psi=(\dot \mu,\dot T, \dot\gamma)^{\rm T}$ with the help of
the matrix $\widehat D$ [see Eq.~(\ref{eq:matrices})], i.e.,
$\dot\Phi=\widehat D\dot\Psi$.  We then obtain $\dot\mu$ and $\dot T$
by imposing the condition that the first two components of $D\dot\Psi$
vanish.  The remaining component $\dot\vartheta$ then takes the value
\begin{equation}
  \label{eq:dot_theta}
  \dot\vartheta=\frac{4 \pi ^4 T^4}{45 h m^2 v_F^5}\dot\gamma,
\end{equation}
where we used Eq.~(\ref{eq:D}) for $\widehat D$.  It has the meaning
of the rate of change of the density of $\Theta$ due to the scattering
processes violating conservation of momentum.

Next, we add $\dot\vartheta$ given by Eq.~(\ref{eq:dot_theta}) with
$\dot\gamma$ given by Eq.~(\ref{eq:dot_gamma}) to the right-hand side
of Eq.~(\ref{eq:continuity_theta}) and solve the system of equations
(\ref{eq:continuity_equations}) in linear order in small $\delta\mu$,
$\delta T$, and $\delta\vartheta$ introduced via
Eq.~(\ref{eq:plain_wave}).  In the low-temperature regime $T\ll
mv_F^2$, the frequency $\omega$ for a given $q$ can be presented as
\begin{equation}
  \label{eq:exact_form}
  \omega=v_Fq\left(1+\frac{T}{mv_F^2}x\right),
\end{equation}
where $x$ satisfies
\begin{equation}
  \label{eq:cubic_equation}
  x^3+\frac{ix^2}{\eta }-\frac{7 \pi ^2 x}{5}-\frac{i \pi ^2}{3 \eta
  }=0,
  \quad
  \eta=\frac{qT}{mv_F}\tau_P^{}.
\end{equation}
This cubic equation has three solutions $x(\eta)$ that correspond to
the three sound modes.  Although analytic expressions for these
solutions can be obtained, their form is overly complicated.  Thus, we
focus on the important limiting cases of $\eta\to0$ and
$\eta\to\infty$.

One of the three solutions is purely imaginary.  In the limiting cases
we find $x=-i/\eta$ at $\eta\to0$ and $x=-(5/21)i/\eta$ at
$\eta\to\infty$.  This solution corresponds to the sound mode that
propagates at velocity $v_1=v_F$ at high $q$. The other two solutions
of Eq.~(\ref{eq:cubic_equation}) yield the modes propagating with
velocities $v_\pm$.  For these solutions we find
$x=\pm\pi/\sqrt3-8\pi^2i\eta/15$ at $\eta\to0$ and
$x=\pm\sqrt{7/5}\pi-(8/21)i/\eta$ at $\eta\to\infty$.  Using
these expressions, we obtain the frequencies of the sound modes in the
limits of small and large $q$.  At $q\gg mv_F/T\tau_P^{}$ our results for
$x$ at $\eta\to\infty$ upon substitution into Eq.~(\ref{eq:exact_form}) yield
\begin{equation}
  \label{eq:dissipation_high_omega}
  \omega=v_Fq-\frac{5i}{21\tau_P^{}},
  \quad
  \omega=v_F\left(1\pm\sqrt{\frac75}\frac{\pi T}{mv_F^2}\right) q
  -\frac{8 i}{21 \tau_P^{}}.
\end{equation}
At $\tau_P^{}\to\infty$ Eq.~(\ref{eq:dissipation_high_omega}) recovers
our earlier results for the three sound modes in the absence of
momentum relaxation with the sound velocities given by
Eq.~(\ref{eq:velocities_3}).  At finite $\tau_P^{}$ the frequencies of
the sound modes acquire imaginary parts of the order of
$-\tau_P^{-1}$.  The latter are the rates of attenuation of sound due
to the momentum relaxation processes.

Our expressions for $x$ at $\eta\to0$ yield the sound mode frequencies at
$q\ll mv_F/T\tau_P^{}$,
\begin{equation}
  \label{eq:dissipation_low_omega}
  \omega=v_Fq-\frac{i}{\tau_P^{}},
  \quad
  \omega=v_Fq\left(1\pm \frac{\pi T}{\sqrt{3} m v_F^2}\right)
  -\frac{8 \pi^2i q^2
      T^2\tau_P^{}}{15 m^2 v_F^2}. 
\end{equation}
At $q\to0$ the first mode is overdamped, while the other two
experience only weak attenuation.  Their velocities recover our
earlier result (\ref{eq:velocities_2}).

The effect of weak static disorder on the three sound modes can be
summarized as follows.  At high $q$ each of the three modes is
attenuated with the rate of the order of $\tau_P^{-1}$.  However, their
behavior at lower $q$ is dramatically different.  The sound mode
propagating with velocity $v_1=v_F$ remains attenuated with the rate
of the order of $\tau_P^{-1}$, and at $\omega\lesssim\tau_P^{-1}$ this
sound mode ceases to exist.  On the other hand, the attenuation rates
of the other two modes depend on $q$ in such a way that they remain
weak (i.e., small compared to the real part of $\omega$) at all
frequencies.  The main effect of the momentum relaxation is the
crossover of the sound velocities $v_\pm$ from their values given by
Eq.~(\ref{eq:velocities_3}) at $\omega\tau_P^{}\gg mv_F^2/T$ to
Eq.~(\ref{eq:velocities_2}) at $\omega\tau_P^{}\ll mv_F^2/T$.

It is instructive to compare the behavior of sound modes in chiral
one-dimensional systems with that in nonchiral ones.  In the latter
case the nature of the two sound modes depends on the presence of
spins in the system.  For fermions with spin, the two modes are very
similar to sound modes in superfluid $^4$He \cite{landau_fluid_2013,
  khalatnikov_introduction_2000}.  In particular, they have different
velocities at $T\to0$, the first sound is a wave of particle density,
whereas the second sound is the wave of entropy
\cite{matveev_second_2017}.  For spinless one-dimensional quantum
fluids the two sound modes have the same velocity at $T\to0$.  At
small but finite temperature the degeneracy is split, but the two
resulting modes are hybrids of the first and second sounds: the
amplitudes of the oscillations of density $\delta n$ and entropy
density $\delta s$ are of the same order of magnitude
\cite{matveev_hybrid_2018}.  The two sound modes in the case of chiral
fermions with disorder are of a similar hybrid nature, with both
velocities approaching $v_F$ at $T\to0$ [see
Eq.~(\ref{eq:velocities_2})] and $\delta s/\delta n=\pm\pi/\sqrt3$.
In the absence of disorder, the modes propagating with velocities
$v_\pm$ are also hybrids, with $\delta s/\delta n=\pm\pi\sqrt{7/5}$,
whereas the additional mode propagating at velocity $v_F$ is a wave of
density, analogous to the first sound: $\delta s/\delta n\to0$ at
$T\to0$.  Similarly to the ordinary sound, this mode is effectively
damped at small $q$ in the presence of disorder.  On the other hand,
disorder results in only weak attenuation of the two hybrid modes.

It is important to distinguish between the sound modes discussed in
this paper and the bosonic excitations in the Luttinger liquid theory
\cite{haldane_luttinger_1981}.  Indeed, our model of chiral spinless
fermions can be bosonized, resulting in an alternative theory based on
bosonic rather than fermionic elementary excitations.  The bosons are
essentially quantized waves of particle density that propagate at the
Fermi velocity.  The nonlinear corrections to the energy spectrum
(\ref{eq:energy_spectrum}) translate into anharmonic coupling of
bosons \cite{haldane_luttinger_1981}, resulting in a finite lifetime
$\tau_b\sim\hbar mv_F^2/T^2$ for excitations with energy of order $T$
\cite{samokhin_lifetime_1998}.  The lifetime $\tau_b$ is usually
shorter than the lifetime of fermionic quasiparticles $\tau_{\rm ex}$
\cite{ristivojevic_relaxation_2013, protopopov_equilibration_2015}.
At time scales beyond $\tau_{\rm ex}$, neither type of quasiparticles
remains well defined, and the hydrodynamic description of the fluid
presented here becomes appropriate.

The two sound modes in the disordered system can be studied in an
experiment with quantum Hall devices, where propagation of a density
pulse along the edge is observed.  At time scales $t\gg\tau_{\rm ex}$
the pulse is expected to split in two, propagating with velocities
$v_+$ and $v_-$ given by Eq.~(\ref{eq:velocities_2}).  An experiment
of this type was reported in Ref.~\cite{ashoori_edge_1992}, where the
density pulse was created by a local gate.  The propagation of the
density pulse was observed, but it did not split into two pulses.  A
possible reason is that at the very low measurement temperature of
0.3K the relaxation time $\tau_{\rm ex}$ may have exceeded the time
during which the pulse was observed.  For spinless chiral fermions
$\tau_{\rm ex}$ scales with the temperature as $ T^{-2}$ and $T^{-6}$
\cite{ristivojevic_relaxation_2013} for pure Coulomb interactions and
those screened by a nearby gate, respectively, and as $T^{-14}$ in the
case of very short range interactions
\cite{protopopov_equilibration_2015}.  Thus the conditions for the
observation of sound modes are more favorable at higher temperatures.
We mention finally that in the case of a disorder-free system, only
the first sound mode is efficiently coupled to the gate, because to
leading order the oscillations of the chemical potential in the hybrid
modes vanish, see Eq.~(\ref{eq:deltas_3_pm}).

To summarize, we have studied sound modes in a system of chiral
one-dimensional weakly interacting fermions.  In the absence of
disorder the system supports three sound modes: the first sound
propagating at the Fermi velocity and
two hybrid modes with velocities $v_\pm$ different from $v_F$ by
$\delta v\sim\pm T/mv_F$.  Disorder effectively damps the first
sound, but the hybrid modes propagate with little damping.

\begin{acknowledgments}

  The author is grateful to A. V. Andreev for helpful discussions.
  This work was supported by the U.S. Department of Energy, Office of
  Science, Basic Energy Sciences, Materials Sciences and Engineering
  Division.
  
\end{acknowledgments}

\bibliography{chiral-sound}
\end{document}